\documentclass[%
 reprint,
 amsmath,amssymb,
 aps,
floatfix,
]{revtex4-1}
\makeatletter
\usepackage{subcaption}
\usepackage[utf8]{inputenc}

\newcommand{\Rmnum}[1]{\expandafter\@slowromancap\romannumeral #1@}
\makeatother
\usepackage{graphicx}
\usepackage{dcolumn}
\usepackage{float}
\usepackage{amsmath}
\usepackage{amsfonts}
\usepackage{amssymb}
\usepackage{color}
\usepackage{units}
\usepackage{verbatim}
\usepackage{bbold}
\usepackage{soul}

\newcommand{\beq}{\begin{equation}}
\newcommand{\eeq}{\end{equation}}
\newcommand{\bea}{\begin{eqnarray}}
\newcommand{\eea}{\end{eqnarray}}


\begin{document}

\title{Optimal Coherent Filtering for Single Noisy Photons}

\author{S. Gao$^1$}
\author{O. Lazo-Arjona$^1$}
\author{B. Brecht$^{1,2}$}
\author{K. T. Kaczmarek$^{1,3}$}
\author{S. E. Thomas$^{1,4}$}
\author{J. Nunn$^5$}
\author{P. M. Ledingham$^1$}
\author{D. J. Saunders$^{1*}$}
\author{I. A. Walmsley$^1$}
 \email{dylan.saunders@physics.ox.ac.uk,\\*
 ian.walmsley@physics.ox.ac.uk}
\affiliation{%
 $^1$Clarendon Laboratory, University of Oxford, Parks Road, Oxford OX1 3PU, United Kingdom\\
 $^2$Integrated Quantum Optics, Universit\"at Paderborn, Warburger Strasse 100, 33098 Paderborn, Germany\\
 $^3$Groupe de Physique Appliquée, Universit\'e de Gen\`eve, CH-1211, Gen\`eve, Switzerland\\
 $^4$QOLS, Blackett Laboratory, Imperial College London, London SW7 2BW, United Kingdom\\
 $^5$Centre for Photonics and Photonic Materials, Department of Physics, University of Bath, Claverton Down, Bath BA2 7AY, United Kingdom}%

\date{\today}

\begin{abstract}

We introduce a filter using a noise-free quantum buffer with large optical bandwidth that can both filter temporal-spectral modes, as well as inter-convert them and change their frequency. We show that such quantum buffers optimally filter out temporal-spectral noise; producing identical single-photons from many distinguishable noisy single-photon sources with the minimum required reduction in brightness. We then experimentally demonstrate a noise-free quantum buffer in a warm atomic system that is well matched to quantum dots and can outperform all intensity (incoherent) filtering schemes for increasing indistinguishability.  

\end{abstract}

\maketitle

Single photons are required for many quantum technologies, for example, quantum communication~\cite{Krenn2016Quantum}, quantum metrology~\cite{G2015Quantum}, optical quantum computating~\cite{O'Brien1567}, and quantum networks~\cite{Kimble2008}. There are three key metrics to characterise single-photon sources: 1) single-photon purity, characterized by the second-order intensity autocorrelation function $g^{(2)}(\tau=0)$, where $\tau$ is the time delay between two arms of an Hanbury Brown-Twiss interferometer;  2) brightness ($B$), which is the probability of having one photon per trial, defined at or after collection optics; 3) indistinguishability. In this paper, we define two kinds of indistinguishablity: a) self-indistinguishalibity $I^{(1)}$, describing the Hong-Ou-Mandel (HOM) dip visibility of photons coming from the same photon source generated at different times; b) inter-indistinguishability $I^{(2)}$, refering to the HOM dip visibility of photons from different sources. $I^{(1)}$ and $I^{(2)}$ are determined by the overlap of photon wavefunction in all degrees of freedom. A perfect single-photon source is characterized by: $g^{(2)}(0)=0$, $B=1$, $I^{(1)}=1$ and $I^{(2)}=1$.

\begin{figure}
\begin{center}
\includegraphics[width=.9\linewidth]{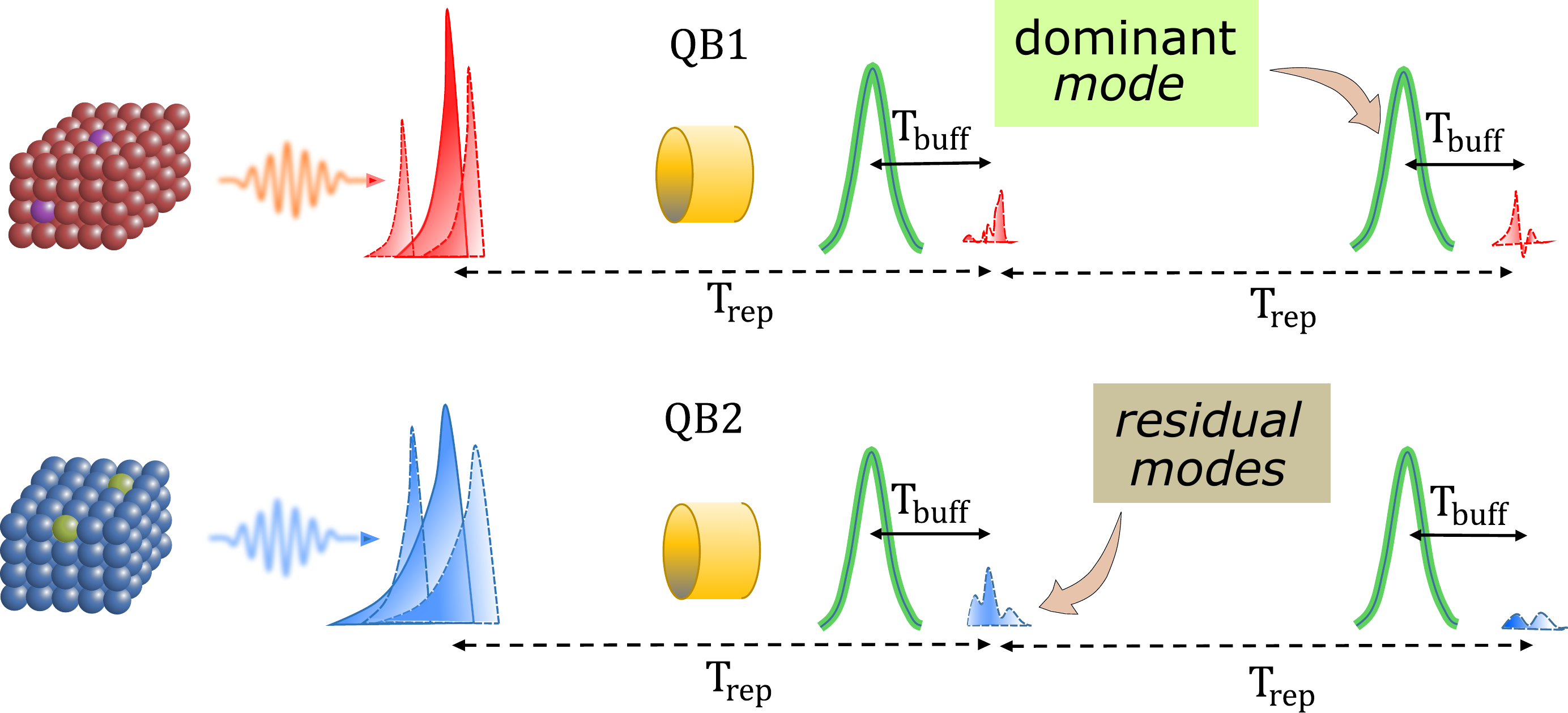}
\end{center}
\vspace{-3ex}
\caption{
Two single-photon sources emit photons that are incoherently mixed ($I^{(1)}<1$) and in different temporal-spectral modes ($I^{(2)}<1$). After filtering with the quantum buffers (QB), the main mode and the residual modes are separated by a short programmable delay $T_{buff}$, pure ($I^{(1)}=1$) and identical photons ($I^{(2)}=1$) are recalled by filtering out the residual noisy modes.}
\label{fig:QD and QM}
\vspace{-3ex}
\end{figure}

Recently, steady progress has been made for single-photon sources with near ideal single photon statistic~\cite{Senellart2017,Schweickert2018a}. The main remaining factor that limits the indistinguishability is the mixing of temporal-spectral optical modes, because of numerous additional undesired physical processes (see review~\cite{Eisaman2011Invited}) in most single-photon emitters such as defects in the solid state~\cite{Aharonovich2016}. To circumvent this issue, one solution is to filter the photons after emission. However, passive intensity filtering only achieves $I^{(1)},I^{(2)}\rightarrow1$ in the limit of $B\rightarrow 0$~\cite{PhysRevLett.100.133601} because the noise spectrum is not intensity-orthogonal to the signal. In this paper, we propose a coherent filter, a quantum buffer, which can increase both $I^{(1)}$ and $I^{(2)}$ to unity whilst minimizing the decrease in brightness. The basic principle of how a quantum buffer increases indistinguishability is shown in Fig. \ref{fig:QD and QM}. A buffer is placed after each single-photon source to delay a user-specified temporal-spectral wavepacket, or temporal-mode~\cite{PhysRevX.5.041017}. The output of all buffers will be pure ($I^{(1)}=1$) and identical ($I^{(2)}=1$). The brightness is optimal if we delay the dominant mode and filter away only the the unwanted residual modes. In order to investigate the feasibility of our scheme, we perform a proof-of-principle experiment that is well matched to state-of-art quantum dot single-photon sources~\cite{Somaschi2016}.

To begin, we consider the state $\rho$ which represents the temporal-spectral degrees of freedom of an imperfect single-photon source. Quantum buffering works by coherently filtering out the largest eigenmode $|\psi_0\rangle$ of $\rho$, where we decompose $\rho=\sum \alpha_k|\psi_k\rangle\langle \psi_k|$, with eigenvalues $\alpha_k$ and eigenmodes $|\psi_k\rangle$~\cite{PhysRev.130.2529}. All the other eigenmodes of the emitted photon are not absorbed by the quantum buffer and are separated in time from the dominant mode and can be removed. As a result, the output of our quantum buffer will be in a temporal-spectral pure state, $\rho_{\text{out}}=|\phi\rangle\langle \phi|$. If all the other degrees of freedom of the emission are identical and the second-order correlation function $g^2(0)$ of the buffered photons remains at $0$, $I^{(1)}=1$ would be achieved. An ideal quantum buffer selects the dominant eigenmode with unit efficiency, and therefore the brightness will decrease by the largest eigenvalue, $B_{\textrm{out}}=\alpha_0 B_0$, where $B_0$ is the brightness of the single-photon source before the quantum buffer. Since different sources may have different dominant temporal modes $|\psi_0\rangle$, the filtered mode will differ from source to source. An ideal quantum buffer will unify these modes, producing the same output mode $|\phi\rangle$ for every single-photon source, making both $I^{(1)}$ and $I^{(2)}$ unity.

\begin{figure}
    \centering
    \begin{subfigure}[b]{.5\textwidth}
        \includegraphics[width = .9\textwidth]{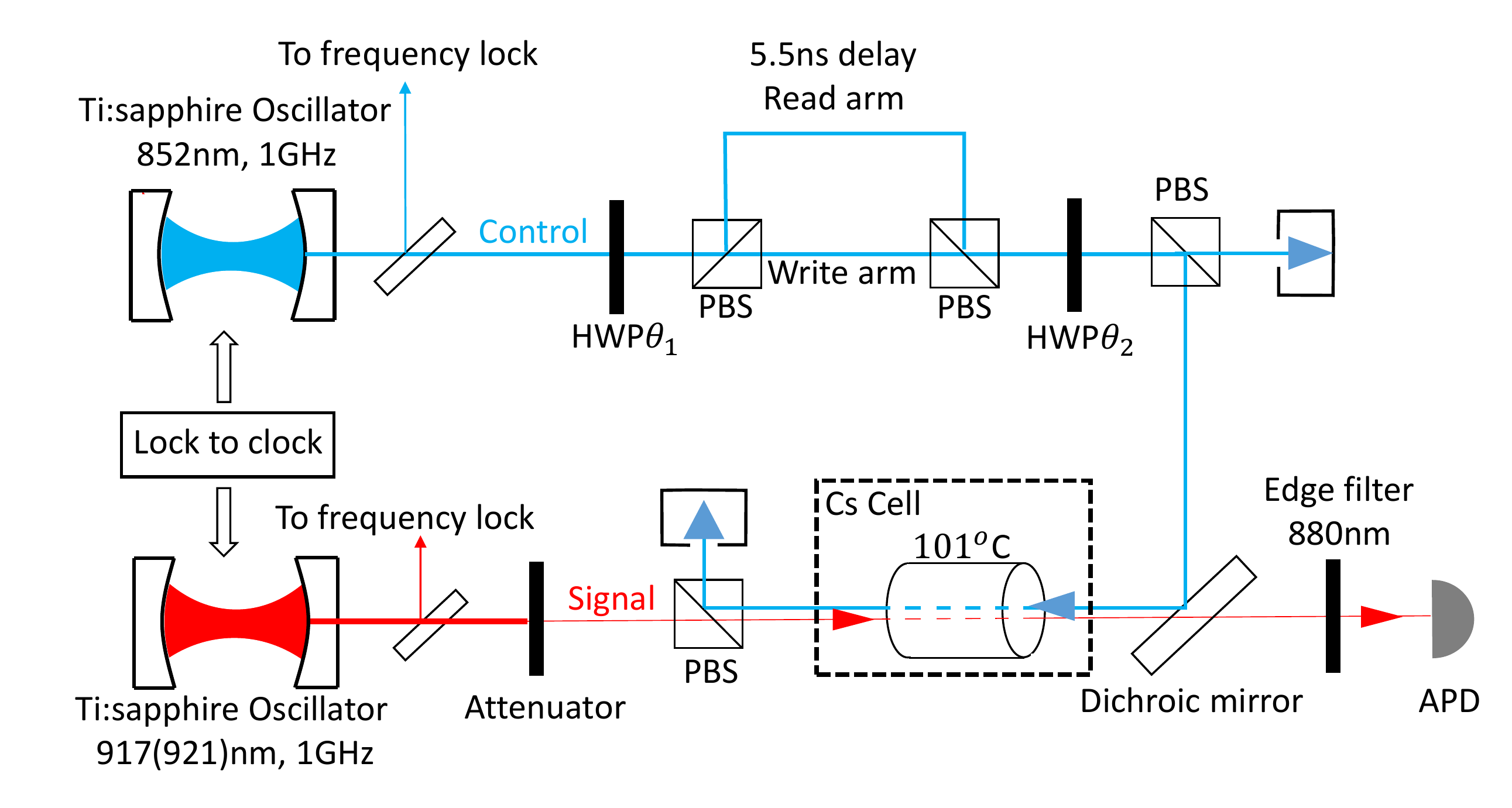}
        \caption{}
        \label{fig:experiment}
    \end{subfigure}
   
    \begin{subfigure}[b]{.17\textwidth}       
        \includegraphics[width=\textwidth]{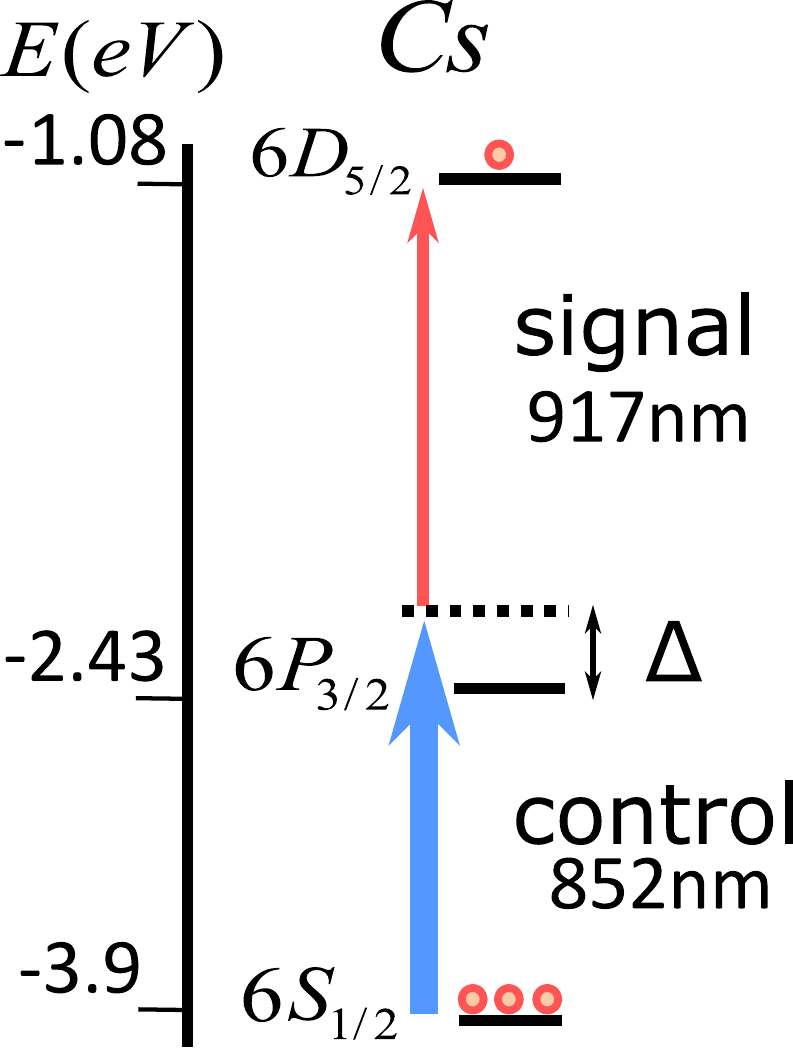}
        \caption{}
        \label{fig:Cs}
    \end{subfigure}
    ~
    \begin{subfigure}[b]{.29\textwidth}
        \includegraphics[width = \textwidth]{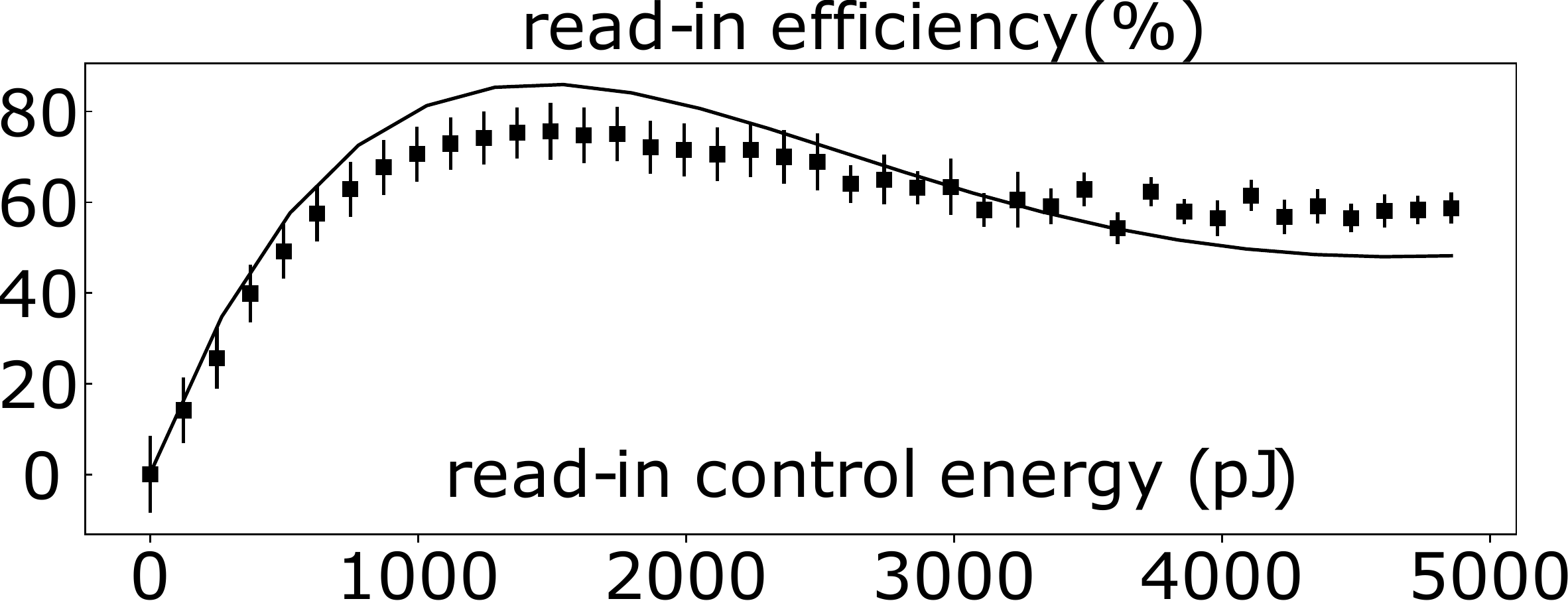}
        \includegraphics[width = \textwidth]{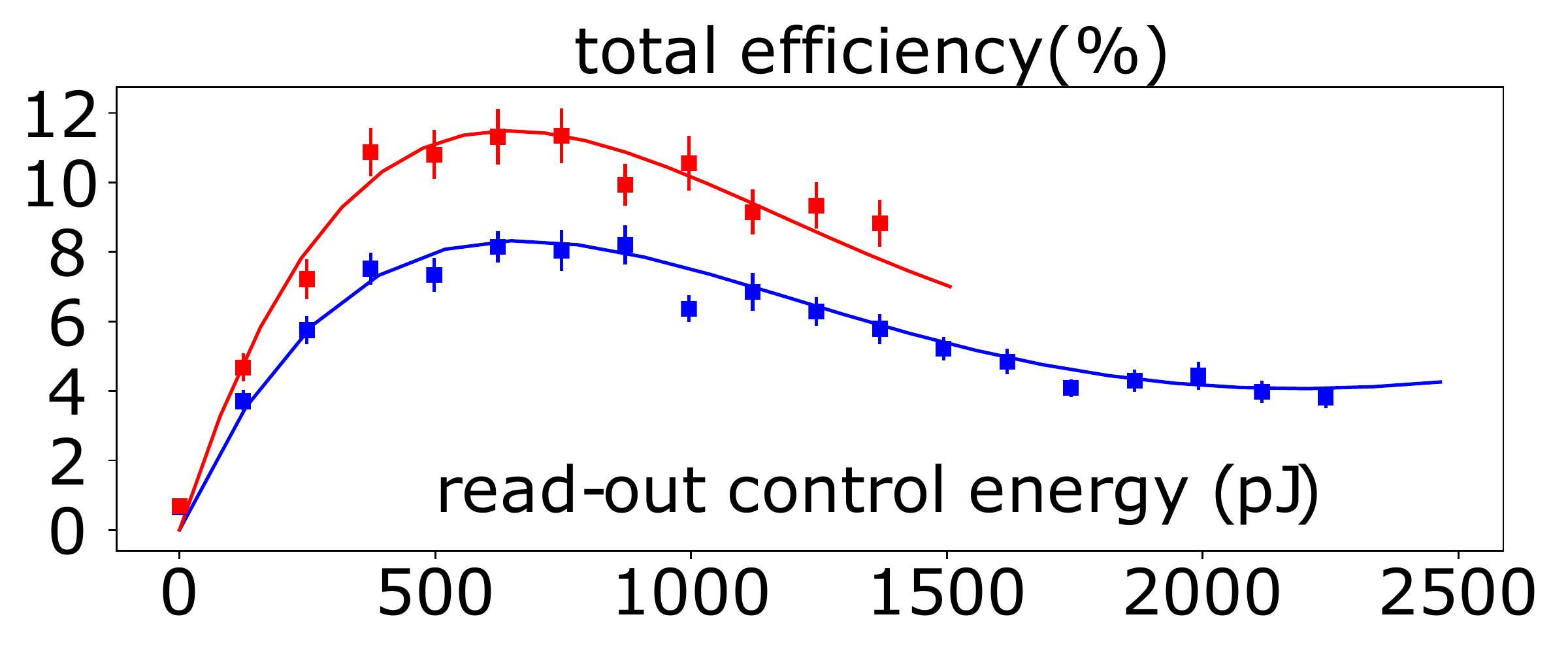}
        \caption{}
        \label{fig:efficiency}
    \end{subfigure}    
\caption{Experimental demonstration of the ORCA buffer, (a) experimental setup. (b) atomic transition of an ORCA Buffer. $\Delta$ is the frequency detuning from single-photon resonance. (c) buffer efficiency for different read-in and read-out control pulse power. Solid line is given by the theory in supplementary materials \Rmnum 2 \cite{supp}. square dots with error bars are experimental data. We show total efficiency for different read-in control powers in the lower figure: blue for $700$pJ and red for $1310$pJ.}
\label{fig:ORCA Buffer}
\end{figure}

Devices that can select particular temporal modes have been demonstrated previously~\cite{Eckstein:11,Ansari:18,Reddy:18,PhysRevX.7.031012,Shahverdi2017Quantum}. However, their bandwidth is limited due to phase-matching constraints and is not matched to leading solid-state single photon sources and careful engineering of the underlying nonlinear process is required to suppress spurious noise processes and render noise levels quantum compatible. To construct a noise-free broadband quantum buffer we implement an Off-Resonant-Cascaded-Absorption (ORCA) buffer; closely related to our recent demonstration of a noise-free ORCA memory in warm caesium (Cs) vapour~\cite{Kaczmarek2017}. The ORCA protocol is based on reversible coherent off-resonant two-photon absorption. Here in our current experiment, the control field is off-resonantly applied to the D-line ($S_{1/2}\rightarrow P_{3/2}$) resonance whereas the signal is applied to the $P_{3/2}\rightarrow D_{5/2}$ (Fig. \ref{fig:ORCA Buffer}b). This is the reverse configuration of the original ORCA memory. This provides access to a larger range of operating wavelengths, (see Fig. \ref{fig:ORCA Buffer}\subref{fig:Cs}). Of particular interest are the 917 and 921nm transitions in Cs that are well matched to many state-of-the-art quantum dots (QDs) single-photon sources, e.g.~\cite{Senellart2017}. In this configuration, an input state, such as a QD-emitted photon, is stored into an atomic coherence shared between the ground and doubly-excited state, mediated by a strong control field. This atomic coherence is then reversibly read-out of the buffer on-demond back into an optical signal by application of a second control field after a short buffering time. We verify that the equations of motion of the new ORCA buffer are the same as that of the ORCA memory (see supplementary materials \Rmnum 2 \cite{supp}), suggesting that all the benefits of the ORCA protocol remain, including broadband and zero-noise operation. The bandwidth of the ORCA buffer is set by the bandwidth of the control field, and can easily accommodate the GHz bandwidth of many single-photon sources~\cite{Senellart2017}. To pinpoint the exact frequency, we can adjust the detuning of the control field, such that the two-photon resonance condition matches the source emission.

To demonstrate the compatibility of the ORCA buffer with single-photon sources we implement a proof-of-principle experiment showing that we can selectively store a specific temporal mode and retrieve it with no added noise using weak-coherent states. A 852 nm control field is produced from a 1.5 GHz bandwidth mode-locked Ti:Sapphire laser, while a signal pulse  is generated using a second 1.5 GHz bandwidth Ti:saph laser, synchronized to the first by means of Spectra Physics Lok-to-Clock\textsuperscript{\textregistered} electronics. The large wavelength mismatch between signal and control allows for simple filtering of the control field using dichroic interference filters. The experimental layout is shown in Fig. \ref{fig:ORCA Buffer}\subref{fig:experiment}. Our ORCA buffer is implemented in a vapour cell with Cs heated to $\sim101^\circ$C. The signal and control overlap in the cell, are focused to a diameter of 210 $\mu m$, and are counter-propagating to reduce Doppler dephasing~\cite{Kaczmarek2017}. The control and signal are tuned 7.5 GHz off resonance with the intermediate state, reducing linear absorption. With a Gaussian shaped control pulse and a buffer delay of 5.5 ns, we observe total efficiencies $\eta_{\textrm{out}}\sim$10\%, and noise-free performance with $\mu_1<10^{-4}$ (where $\mu_1 =$ noise/efficiency is a common measure of such performance~\cite{PhysRevLett.114.230501}). The small amount of residual noise is comparable with observed detector dark count rates confirming that our Cs ORCA buffer will not affect the $g^{(2)}(0)$ of any stored light, as observed in our ORCA memory~\cite{Kaczmarek2017}.

We measured the memory performance as a function of read-in and read-out control field powers, and developed a model that gives good agreement with the data, as seen Fig~\ref{fig:ORCA Buffer}\subref{fig:efficiency} (see supplementary materials \Rmnum 3 \cite{supp}). The roll over of efficiency is due to secondary read-out and read-in of the signal during retreival, and this process is an interplay of many parameters such as control pulse power, number density of the vapour and the cell length. By factoring out Doppler dephasing~\cite{Zhao2009}, which is avoided at the very short memory times required for quantum buffering we predict a maximum efficiency for our current experiment of 34\%, which is limited by current control pulse power. 

Using our theoretical model, we now numerically predict the filtering performance of our current Cs ORCA buffer implementation for filtering solid-state single-photon source emission. The indistinguishability and the brightness are both determined by the \emph{singlemodeness} of our ORCA buffer. To explore this, we consider the linear map of the input optical mode to the output optical mode, as described by the Green's function (see supplementary materials \Rmnum 2 \cite{supp}:
\begin{equation}
\label{eq:Green}
S_{\textrm{out}}(t) = \int G(t,t')S_{\textrm{in}}(t')dt'.
\end{equation}
A singular value decomposition (SVD) of the Green's function ($G(t,t')=\sum_k\lambda_k|\varphi_k\rangle\langle u_k|$) will give a series of orthogonal input signal modes $|u_k\rangle$ mapping to orthogonal output $|\varphi_k \rangle$ with a buffer efficiency equal to the square of the corresponding singular value $\lambda_k$. A single mode memory is a memory with only one non-zero $\lambda_k$, meaning that there is only one input mode $|u_0\rangle$ that can interact with the ORCA buffer. We can engineer the ORCA buffer such that $|u_0\rangle$ matches the dominant mode $|\psi_0\rangle$ of the single-photon source emission $\rho$ by shaping the temporal mode of the control-field. The brightness will then be: $B_{\textrm{out}} = B_0\alpha_0|\lambda_0|^2$, where $B_0$ is the brightness of the photon source prior to our quantum buffer and $\alpha_0$ is the fraction of the dominant mode of the source emission. 

For our ORCA buffer, the device becomes multimode due to the choices of several parameters, such as the length of the atomic medium, the detuning, the control field the temporal shape and its energy~\cite{Oscar2019}. The output of an ORCA buffer is a mixture of different modes described by the density matrix
\beq
\rho_{\textrm{out}}=\frac{1}{W} \sum_k \xi_k \alpha_k|v_k\rangle\langle v_k|,
\eeq
\noindent where $|v_k(t)\rangle = \int G(t,t')|\psi_k(t')\rangle dt'$ is the output for input mode $|\psi_k\rangle$, $\xi_k$ is the buffer efficiency for the optical mode $|\psi_k\rangle \rightarrow |v_k\rangle$ and $W=\sum_k \xi_k \alpha_k$ is the normalization constant. The brightness after the buffer is $B=W$; the indistinguishability of the output photon will be $I_{\textrm{out}}=1/K_{\textrm{out}} = Tr[\rho_{\textrm{out}}^2] $ where $K$ is the Schmidt number~\cite{Grobe1994}.

To investigate the applicability of our buffer we turn our attention to its application to GHz bandwidth quantum dot emission~\cite{Senellart2017,PhysRevLett.118.253602}. Semiconductor quantum dots are a leading single-photon source candidate, with unprecedented brightness and very low $g^{(2)}(0)$~\cite{Gammon1996Homogeneous,Sebald2002Single,Couteau2004Correlated,Holmes2014Room,Schweickert2018a}. However, QDs (and some other solid-state sources~\cite{Aharonovich2016}) suffer from fundamental local environmental fluctuations fast enough to contribute to pure dephasing, which limit the indistinguishability, $I^{(1)}<1$ even at $0$ K~\cite{PhysRevB.90.035312,Loredo:16}. To suppress these dynamics as well as to direct the emission, QDs are typically embedded in waveguides~\cite{Claudon2010A, Arcari2014} or micro cavities~\cite{Somaschi2016,Ding2016On}. However, there is a trade-off between brightness and indistinguishability~\cite{Jake2017Phonon}. 
Our quantum buffer is well suited to solve these issues as it is bandwidth and wavelength matched, and can also operate at the required high repetition rates as it has no need for timely state preparation.

We start by considering a single photon emitted by a QD. Its temporal-spectrum mode is described by a two-colour spectrum $C(\omega,\mu)_{\omega_0,t_0} = \langle E^+(\omega) E(\mu)\rangle_{\omega_0,t_0} $ where $E^+$ and $E$ are the creation and annihilation operators of the electric field. $\omega_0$ and $t_0$ are the central frequency and the emission time. This two-colour spectrum has only pure dephasing dynamics which sets the ultimate fundamental limit of indistinguishability of QD emission~\cite{Kaer2014,Jake2017Phonon,PhysRevB.97.195432,PhysRevB.95.201305}. The emission further decoheres due to frequency diffusion~\cite{Kuhlmann2013Charge,PhysRevB.63.155307,PhysRevB.68.233301,PhysRevB.69.041307} and temporal jitter which are induced by both the host material and excitation scheme. The general form of the two-colour spectrum of the emitted photons will be 
\begin{equation}
c(\omega,\mu) = \int p(t_0,\omega_0)C(\omega,\mu)_{\omega_0,t_0}dt_0 d\omega_0
\end{equation}
where $p(t_0,\omega_0)$ is a probability distribution capturing such inhomogeneous broadening. The photon density matrix in frequency representation is the normalized two-color spectrum in matrix form $\rho_{\omega\mu} = c(\omega,\mu)$. Unit indistinguishability is only possible if $\rho$ represents a pure state (see supplementary materials \Rmnum 4 \cite{supp}).

\begin{figure}[t!]
\centering
    \begin{subfigure}[b]{.5\textwidth}
    \includegraphics[width = .9\textwidth]{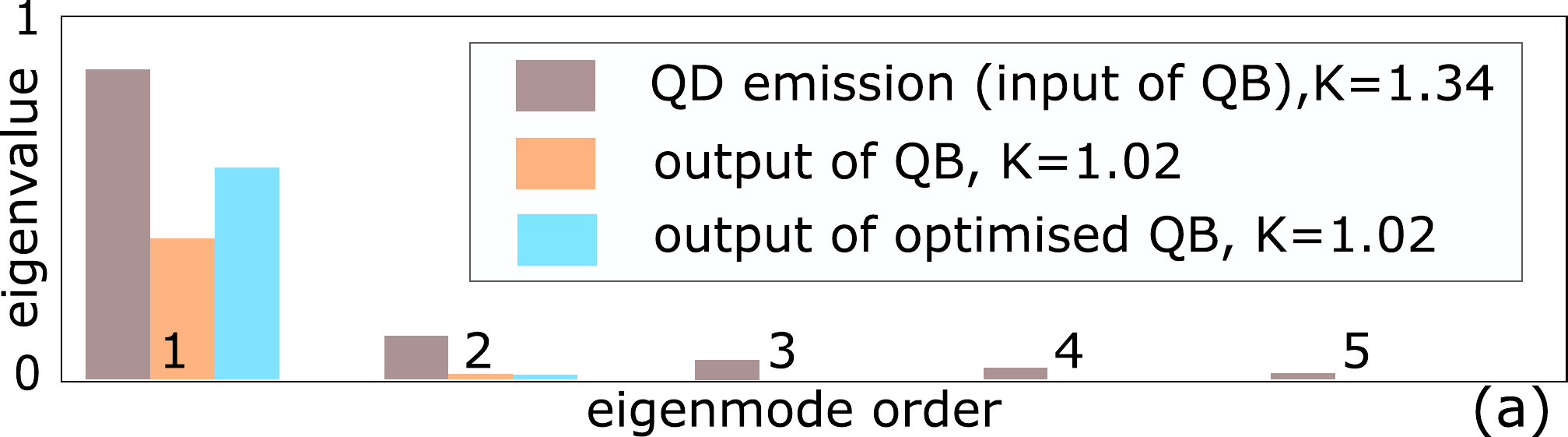}
    \label{fig:singlemodeness}
    \end{subfigure}
    
    \begin{subfigure}[b]{.5\textwidth}
    \includegraphics[width = .9\textwidth]{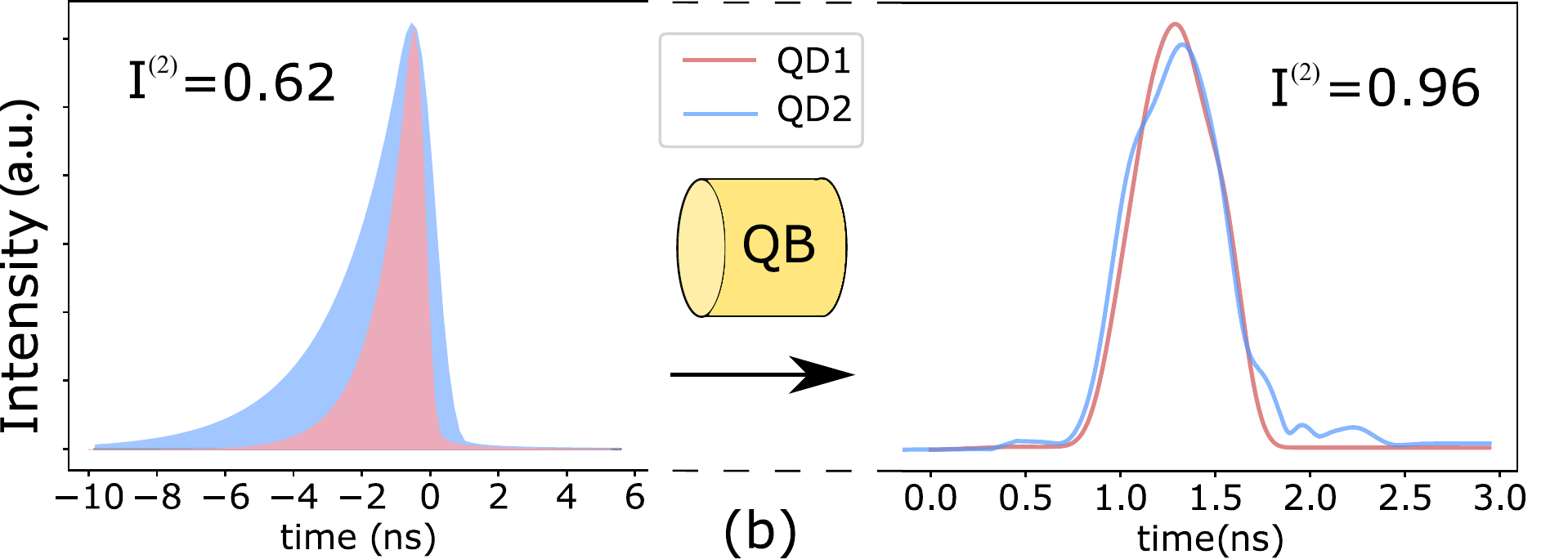}
    \label{fig:mode conversion}
    \end{subfigure}
    
\caption{(a) Demonstration of singlemodeness of an ORCA buffer. The y-axis is the fractions of the first five optical modes before (brown) and after (orange and blue) the ORCA buffer. $K$ is the Schmidt number (mode number). (b) Mode unification by an ORCA buffer. Emission from two distinct QDs ($I^{(2)}$ = $0.62$) with different decay times (left figure) are actively unified by an ORCA buffer into two nearly identical photons (right figure, $I^{(2)}$ = $0.96$).}
\label{fig:single mode and conversion}
\end{figure}

To demonstrate coherent filtering using the ORCA buffer we consider a few examples. We simulate a state-of-the-art off-resonantly excited QD with a self-indistinguishability of $I^{(1)}\approx 0.7$ (similar to QD3 in Ref.~\cite{Somaschi2016}), and predict the improvement with our Cs ORCA buffer using the same parameters as our experiments described before using pulsed weak coherent light. Here the distinguishability is mainly caused by the timing jitter introduced by off-resonant pumping scheme. Figure \ref{fig:single mode and conversion}\subref{fig:singlemodeness} shows the expected filtering performance. We predict an increase of the indistinguishability to $I^{(1)}_{\textrm{out}}\approx0.98$, with a brightness of $B_{\textrm{out}}=0.4 B_0 \approx 0.29$, where $B_0 = 0.72$ is the initial brightness of the QD. This predicted performance already matches the best resonantly pumped QDs (e.g. QD4 in reference~\cite{Somaschi2016}). To improve performance even further, we numerically optimize the shape and energy of both the read-in and read-out control pulse. The brightness and self-indistinguishability after buffering then is: $I_{\textrm{out}}^{\textrm{(1)}}\approx0.98$ \& $B_{\textrm{out}}\approx0.61B_0\approx 0.43$, see Fig~\ref{fig:single mode and conversion}a.

An important measure of the buffer performance is whether it can do better in preparing pure single photons than passive spectral filtering (i.e. using a time-stationary linear filter). We therefore compare our predicted ORCA temporal-spectral filtering against this conventional passive intensity filtering approach~\cite{He2013On,Wei2014Deterministic} for world-leading quantum dots.

\begin{figure}[t!]
\includegraphics[width = .4\textwidth]{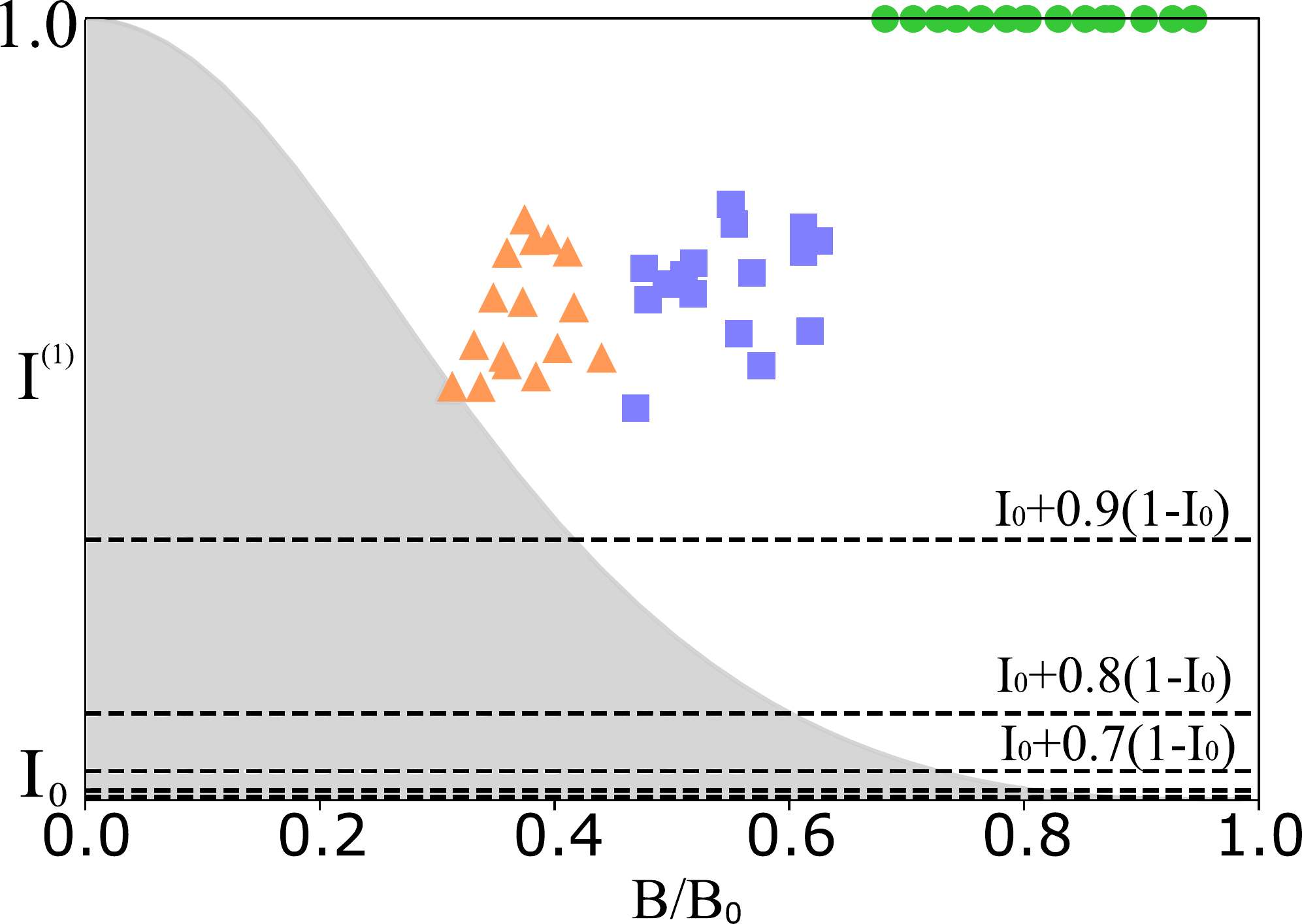}
\caption{Simulation of comparison of purification effect by intensity filtering (shaded area). We plot the brightness B normalized by the initial brightness $B_0$, against the self-indistinguishability. $I_0$ is the initial self-indistinguishability. Dashed-line is $I^{(1)} = I_0+q(1-I_0)$ and $q$ is a fraction number indicating the improvement of $I^{(1)}$. The shaded region is the possible performance of intensity filtering. The triangles indicate ORCA buffer filtering with Gaussian temporal-mode control pulse for 16 random samples. The squares are ORCA buffer filtering with optimized control pulse shape and energy for the same set. The circles show the performance of an ideal quantum buffer.}
\label{fig:result}
\end{figure}

To explore this landscape we simulate 16 different QDs with various pure dephasing magnitudes, spectral diffusion and timing jitter (see supplementary material \Rmnum 4 \cite{supp} for parameters details). The performance of passive intensity filtering is shown in Fig.\ref{fig:result}, where unit indistinguishability can only be achieved in the limit that the brightness goes to zero. This is because the modes of the QD all overlap spectrally and temporally. Therefore only by a careful choice of spectral passband center and infinitely narrow passband is it possible to discriminate between them. The upper-bound of the intensity filter region is found by calculating 100 QDs with various noise dynamics (see supplementary material \Rmnum 5 \cite{supp} for details about intensity filter model). We also plot the predicted performance of our experimental demonstration of the ORCA buffer system, both with and without control-pulse shape optimization; both outperforming passive filtering. An ideal filter would have $I^{(1)}=1$, and $B=B_0\alpha_0$. We have shown that the equations of motion for the ORCA buffer allow for unit memory efficiencies, with $K=1$~\cite{Oscar2019}, which with proper mode matching can optimally filter QD emission. To improve the predicted performance closer to an ideal quantum buffer additional numerical optimizations are required, for example adjusting the interaction length and temperature, or implementing our ORCA buffer in a low-finesse cavity~\cite{Nunn2017}.
 
Besides improving self-indistinguishability $I^{(1)}$, our quantum buffer can also  convert the output temporal-spectral mode of our buffer (by changing the shape of the read-out control field) to make remote QD emissions identical. We demonstrate this capability numerically by modeling two distinct QD emissions with the same central frequency. They start with a inter-indistinguishability of $I^{(2)}=0.62$, as in Fig~\ref{fig:single mode and conversion}\subref{fig:mode conversion}. We then simulate two ORCA buffers, one interacting with each QD. By selecting the largest eigenmode of each QD photons in their respective ORCA buffers, and then recalling them with appropriately adjusted control pulses we increase the inter-indistinguishability to $I^{(2)}=0.96$ by reading the stored excitations out into nearly identical modes. In most cases QDs are also distinguishable in their central emission frequency. In the above simulations, we keep the central frequencies the same for the two dots. Nevertheless, it is possible, using ORCA, to adjust the readout frequency of the stored excitation as compared to the input. Our simulations show that we can implement frequency conversion up to 1nm without significant drop in efficiency (see supplementary material \Rmnum 6 \cite{supp}), whilst also implementing coherent temporal-mode filtering and conversion in a single device. 

In conclusion, we have introduced a quantum buffer to optimally filter solid state single-photon emission in order to circumvent the distinguishability between generated photons due to temporal-spectral mode mixing ($I^{(1)}<1$) and source-dependent mode mismatch ($I^{(2)}<1$). As an example, we experimentally demonstrated the key performance criteria for a quantum buffer for QDs in Cs vapour. Our noise-free Cs ORCA buffer is compatible in wavelengths and bandwidths with InGaAs QDs, and will enable different QD emission from remote samples to be quantum buffered into pure and identical single-photon sources with no increase in $g^{(2)}(0)$. The ideas presented here are applicable to any noisy optical state and any photon-source. Importantly, the room-temperature quantum buffer circumvents the limitations of passive frequency filtering, and provides a new route to produce identical single photons from imperfect single-photon sources.

We would like to acknowledge Jake Iles Smith for providing a solver for pure dephasing QDs, Christian Weinzetl for comments on the manuscript, Bryn Bell for useful discussion, Jonas Becker for discussions on noise mechanisms and models for the quantum dots outputs, and Pascale Senellart for stimulating and helpful comments on an early manuscript. I.A.W. acknowledge support from the European Research Council, the UK Engineering and Physical Sciences Research Council (project EP/K034480/1 and the Networked Quantum Information Technology Hub), an ERC Advanced Grant (MOQUACINO), and the Air Force Office of Scientific Research (Grant/Cooperative Agreement FA9550-17-1-0064). PML acknowledges financial support from a European Union Horizon 2020 Research and Innovation Framework Programme Marie Curie individual fellowship, Grant Agreement No. 705278. S.E.T is supported by EPSRC via the Controlled Quantum Dynamics CDT under Grants EP/G037043/1 and EP/L016524/1, S.G is supported from National University of Defence Technology of China and O.L-A is supported from from Consejo Nacional de Ciencia y Tecnología, and Banco de México.

\newcommand{\noop}[1]{}

\end{document}